\pdfoutput=1
\documentclass[11pt,a4paper]{article}
\usepackage[margin=2.5cm]{geometry}
\usepackage{graphicx,amsmath,amssymb,booktabs}
\usepackage{siunitx}
\usepackage{caption}
\usepackage{accsupp}
\usepackage{url}
\usepackage[colorlinks=true,allcolors=blue]{hyperref}
\hypersetup{
  pdftitle={Two-dimensional liquid-like order in a confined population of
            swimming Eudorina colonies},
  pdfauthor={Nicholas Tufillaro},
  pdfsubject={Active matter; collective behaviour of swimming algal colonies},
  pdfkeywords={active matter, collective behaviour, volvocine algae,
               Eudorina, pair correlation function, low Reynolds number,
               boundary accumulation, bond-orientational order,
               two-dimensional melting},
  pdflang={en-US},
  pdfdisplaydoctitle=true
}
\usepackage{authblk}
\title{\bfseries Two-dimensional liquid-like order in a confined population of
swimming \emph{Eudorina} colonies}
\author[1]{Nicholas Tufillaro}
\affil[1]{\small Aqualytics, Corvallis, Oregon 97331, USA\\\texttt{nick@aqualytics.eco}}
\date{October 2026}

\begin{document}
\maketitle

\begin{abstract}
\noindent
Swimming microorganisms accumulate at boundaries, but the spatial statistics of
the resulting layers are rarely measured. We report a population of colonial
green algae, of order \SI{71}{\micro\metre} across, that accumulates
against the rigid upper wall of a closed
\SIrange{1}{2}{\milli\metre} chamber and holds a two-dimensional
non-contacting arrangement for about \SI{105}{\minute}. The pair correlation function has a first shell at
$r/\bar{D} = \num{1.73}$, $g = \num{4.8}$; a hard-disc null at the
same area fraction reaches only $g = \num{2.1}$, so the ordering is
not excluded volume alone. Nearest-neighbour separation is
$1.80 \pm 0.13$ diameters. The bond-orientational order parameter
is $\langle|\psi_6|\rangle = \num{0.456}$ against \num{0.385} for the
null, only \SI{36}{\percent} of interior colonies are six-coordinated, and
the orientational correlation $g_6(r)$ decays over
\num{1.5} diameters with $\eta_6 = \num{1.31}$ against the hexatic
bound of $1/4$: the arrangement is a two-dimensional liquid, not a crystal or
a hexatic. Colonies execute a persistent random walk with reorientation time
\SIrange{1.4}{1.7}{\second} at Reynolds number
$\sim 9\times 10^{-4}$. Breakup begins at \SI{127}{\minute}. Thermal convection
is excluded on Rayleigh-number grounds. Caveats: the length calibration is
nominal rather than metrological, and this is a single population observed
once. All data and code are deposited.
\end{abstract}


\section{Introduction}

Motile microorganisms accumulate at solid boundaries, a consequence of
swimming, steric interaction and hydrodynamic coupling to the wall that is
common to bacteria, spermatozoa and algae. In dense suspensions this
accumulation drives bioconvection, described first for \emph{Euglena} by
Wager~\cite{wager1911} and named by Platt~\cite{platt1961}, with continuum
theory following~\cite{childress1975,pedley1992,hill2005,bees2020}. In dilute
populations near a wall the relevant physics is instead pairwise: Drescher
\emph{et al.}~\cite{drescher2009} showed that two \emph{Volvox} colonies
swimming against a boundary attract one another into stable orbiting states,
analysed subsequently by boundary-element simulation~\cite{ishikawa2020}.

Between these limits lies a regime that is less well characterised: a
population dense enough to interact but too dilute to convect, held against a
boundary long enough for its spatial statistics to be measured. Balasubramanian
reported that \emph{Volvox barberi} forms two-dimensional flocks packing close
to optimally, essentially in contact~\cite{balasubramanian2021}. We report
here the opposite case, in a smaller volvocine colony: a stable arrangement in
which colonies are held apart, with a preferred separation near two diameters
and a pair correlation function characteristic of a two-dimensional liquid.

The measurement is straightforward, and we make no claim that the apparatus is
sophisticated. Its interest lies in the specific numbers, in the null model
against which the ordering is tested, and in a measured lifetime for the
structure.

\section{Material and methods}

\subsection{Sample and chamber}

The sample was collected from Lake Oroville, a reservoir in northern
California, on 29 March 2018. The collection is documented in
Ref.~\cite{video1}. It was not a monoculture. It stood beside a window for two weeks before imaging,
receiving natural daylight and no added nutrients, which enriches for motile
phototactic flagellates. Nothing here should be read as a limnological
statement about the reservoir.

For imaging the sample was held in a closed clear plastic coin capsule
(Air-Tite ``Direct Fit'' H40; internal diameter \SI{40.6}{\milli\metre},
internal depth \SI{3.17}{\milli\metre}) filled to a fluid depth of
\SIrange{1}{2}{\milli\metre}. Both bounding surfaces are therefore rigid and
no-slip, matching the geometry assumed in Ref.~\cite{drescher2009}. The nearest
sidewall lies about \SI{20}{\milli\metre} from the imaged field, some
\num{280} colony diameters, so lateral confinement is negligible.

\subsection{Exclusion of thermal convection}

In a layer of depth $d$ the Rayleigh number is
\begin{equation}
\mathrm{Ra} = \frac{g\alpha\,\Delta T\, d^{3}}{\nu\kappa},
\label{eq:rayleigh}
\end{equation}
with $g\alpha/\nu\kappa \approx \SI{1.4e10}{\per\metre\cubed\per\kelvin}$ for
water near room temperature. Taking $\mathrm{Ra}_c \approx 1100$, onset
requires \SI{79}{\kelvin} across the layer at $d=\SI{1}{\milli\metre}$ and
\SI{10}{\kelvin} at \SI{2}{\milli\metre}. Neither is attainable, and the
thermal diffusion time $d^{2}/\kappa \approx \SI{28}{\second}$ at
\SI{2}{\milli\metre} precludes sustaining such a gradient. The structure
reported below is not a convective one; it is also static rather than
circulating, which bioconvective plumes are not.

\subsection{Imaging and its limitations}
\label{sec:optics}

Imaging used a handheld USB digital microscope (AmScope UTP class): a
\num{2}-megapixel CMOS sensor, \num{1600}$\times$\num{1200} effective pixels,
continuously variable magnification, eight built-in LEDs. Two sequences were
recorded: a continuous \SI{117}{\second} run at native rate for dynamics, and a
\num{20}$\times$ time lapse spanning \SI{153}{\minute} of elapsed time
for structure and lifetime; the latter sequence may be viewed directly in
Ref.~\cite{video2}.

An instrument of this kind costs a small fraction of a research or teaching
compound microscope. That was a deliberate choice rather than a constraint, for
the reasons set out in \S\ref{sec:cost}; the price is paid in metrology.

Use of a very low cost imaging system carries significant limitations. The
quoted magnification is not an optical
quantity: it is defined against an unspecified display and set continuously by
working distance, and with no eyepiece there is no field number from which to
compute a field of view. We adopt \SI{1.55}{\micro\metre} per pixel with
an estimated \SI{20}{\percent} uncertainty, but this was not established with a
stage micrometer and should be regarded as nominal. Second, the recorded video
is \num{1280}$\times$\num{720} while the sensor is
\num{1600}$\times$\num{1200}, so the frames were cropped or rescaled by an
unrecoverable amount; this is an additional error path not covered by that
\SI{20}{\percent}. Third, no numerical aperture is published, so no depth of
field can be computed and the planarity of the layer cannot be established
optically. The empirical argument for near-planarity is that objects elsewhere
in the same frames appear as clearly defocused rings while the analysed cluster
is sharp.

Every dimensionless quantity reported below---$g(r)$, the nearest-neighbour
ratio, the mean-squared displacement exponents, the reorientation time---is
independent of the calibration. Quantities in microns are not.

The native-rate clip proved to be a \SI{15}{\hertz} capture padded into a
\SI{30}{\hertz} container, with every second frame duplicated; frames were
decimated by two before tracking. Analysed without decimation the velocity
autocorrelation acquires a spurious period-two oscillation.

\subsection{Identification}
\label{sec:identification}

The organism is \emph{Eudorina}: a coenobium of \num{16}, \num{32} or \num{64}
biflagellate cells embedded at the periphery of a mucilage matrix, typically
\SIrange{50}{200}{\micro\metre} in overall diameter with individual cells
\SIrange{10}{24}{\micro\metre} across. It swims by coordinated flagellar
beating, is bottom-heavy and therefore gravitactic, and is strongly
phototactic.

The identification was made by direct examination under a compound laboratory
microscope at magnification sufficient to resolve the individual cells of the
coenobium.
This is not an observation the low-cost imaging system used for the
quantitative work could have supplied: at its resolution and depth of field
(\S\ref{sec:optics}) individual cells within a colony are not separable. The
morphological measurements that can be made from the video are reported in
\S\ref{sec:morphology}; they are consistent with \emph{Eudorina} but would not
have established it.

\subsection{Detection, tracking and statistics}

Colonies were segmented on a greenness index $I_g = G-(R+B)/2$, Gaussian
blurred, thresholded at a fixed level, and labelled by connected components;
equivalent diameters follow from component area. Trajectories were built by
greedy nearest-neighbour linking with a maximum inter-frame displacement of
\num{12} pixels.

The pair correlation function was computed over all pairs, with separations
normalised by the mean colony diameter in each frame and the histogram
normalised against the two-dimensional ideal-gas expectation, the tail over
$3.5 < r/\bar{D} < 5.5$ used to set the baseline.

For the null model, the two-dimensional area fraction inside the convex hull of
each cluster was measured on the native-rate sequence, giving
$\phi_{\rm nr} = \num{0.17}$. Hard discs were then
placed by random sequential adsorption in a square box of the same area, with
the same colony-size distribution, frame by frame, and $g(r)$ computed identically. The corresponding figure for
the time-lapse sequence, over which the stability statistics of
\S\ref{sec:stable} are computed, is $\phi_{\rm tl} = \num{0.104}$;
the two differ because the sequences resolve different numbers of small
colonies, and each analysis uses the value measured on its own frames.

\subsection{Bond-orientational order}

To distinguish a two-dimensional liquid from a crystal we computed the local
bond-orientational order parameter
\begin{equation}
\psi_m(i) = \frac{1}{N_i}\sum_{j \in \mathcal{N}_i} e^{\,\mathrm{i}m\theta_{ij}},
\label{eq:psi}
\end{equation}
where $\mathcal{N}_i$ are the Delaunay neighbours of colony $i$, $N_i$ their
number, and $\theta_{ij}$ the angle of the bond from $i$ to $j$. Particles on
the convex hull were excluded, since their neighbour shells are truncated. The
same quantity was computed for the hard-disc null configurations. We also
recorded Delaunay coordination numbers, and computed $\psi_4$ as a
square-symmetry control.

Following the Kosterlitz--Thouless--Halperin--Nelson--Young (KTHNY) description
of two-dimensional order, we further computed the orientational correlation
function
\begin{equation}
g_6(r) = \frac{\langle \psi_6^{*}(0)\,\psi_6(r)\rangle}
              {\langle |\psi_6|^{2}\rangle},
\label{eq:g6}
\end{equation}
binned in units of mean colony diameter and averaged over frames, and fitted it
to both an exponential $A\exp(-r/\xi)$ and a power law $Ar^{-\eta_6}$. The same
pair of measures---a translational correlation length together with the decay
of long-range orientational order---has been used to characterise
order--disorder transitions in other driven two-dimensional patterns, for
example the capillary ripple lattices of Ref.~\cite{tufillaro1989}.

\subsection{Robustness}

The entire pipeline was recomputed with a deliberately different segmentation
(a hue-window signal with an Otsu-selected threshold in place of the fixed
threshold on $G-(R+B)/2$), different frame subsets and three minimum-area cuts.
Results of that comparison are reported in \S\ref{sec:robust}, and the script
is included in the deposit.

\section{Results}

\subsection{A stable, non-contacting two-dimensional arrangement}
\label{sec:stable}

\begin{figure}[t]
\centering\BeginAccSupp{method=escape,ActualText={Six microscope photographs in two rows of three, showing the same region of the sample chamber at 21 minute intervals from 0 to 105 minutes. In each panel roughly twenty to thirty dark green spheres are spaced apart from one another on a pale background. The spacing pattern persists across all six panels although individual spheres move and exchange neighbours.}}%
\includegraphics[width=\textwidth]{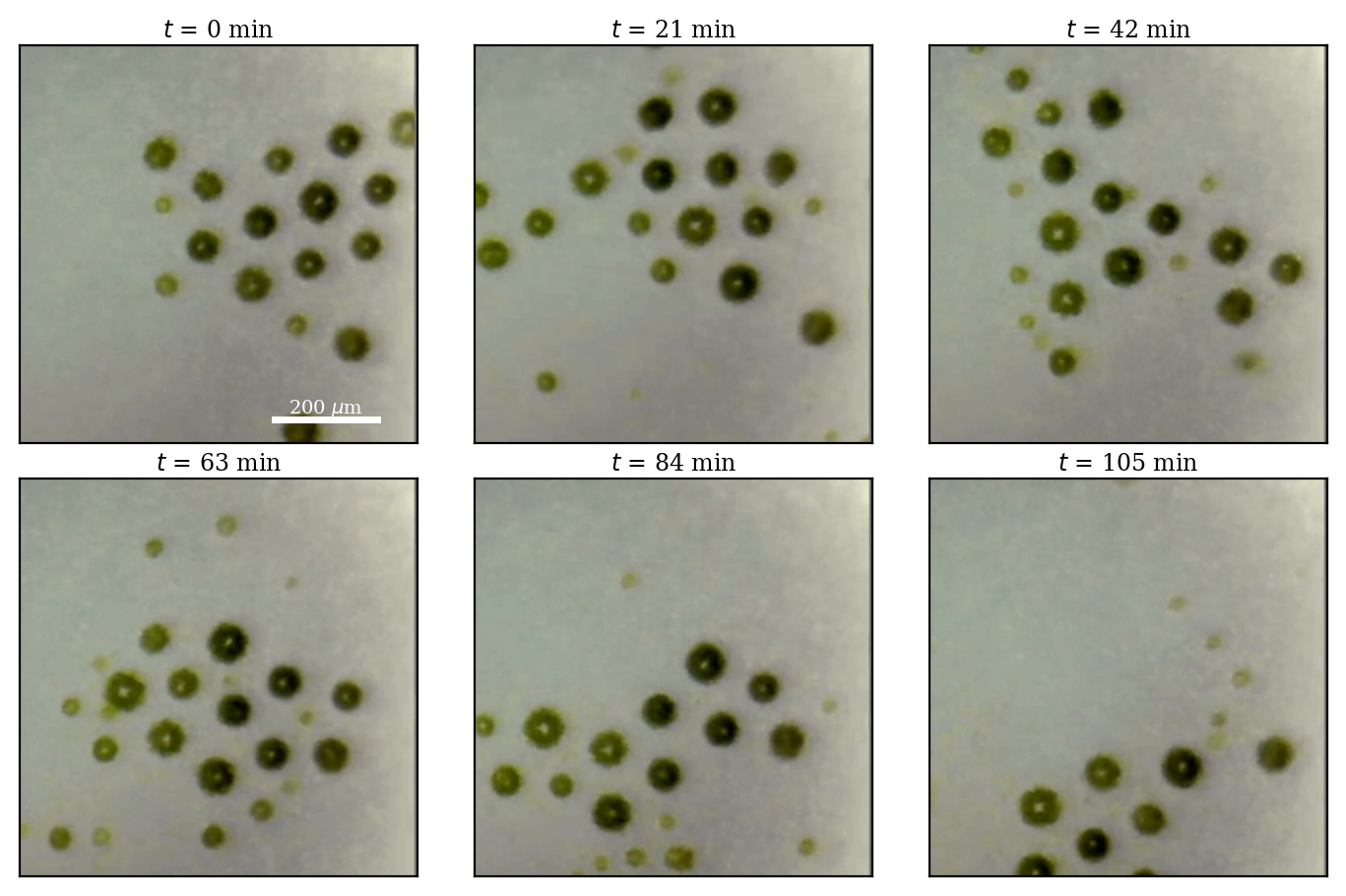}%
\EndAccSupp{}
\caption{Stills from the time-lapse sequence (Ref.~\cite{video2}) at
\SI{21}{\minute} intervals of elapsed time, same region of the field. Colonies exchange neighbours and the
cluster drifts, but characteristic spacing and local coordination persist.}
\label{fig:stills}
\end{figure}

Over the first \SI{105}{\minute} the population holds a loose but clearly
ordered arrangement (figure~\ref{fig:stills}). Median nearest-neighbour
separation, normalised by mean colony diameter, is $1.80 \pm 0.13$ across
\num{93} samples of the record, with a median of \num{30}
colonies in the field and area fraction \num{0.104}. Contact
corresponds to \num{1.0}. Because this is a projection, foreshortening can only
reduce apparent separations, so \num{1.80} is a lower bound on the true
three-dimensional spacing.

\subsection{The ordering exceeds excluded volume}

\begin{figure}[t]
\centering\BeginAccSupp{method=escape,ActualText={Line graph of the pair correlation function against separation in colony diameters. The observed curve is zero below one diameter, rises to a sharp peak of 4.8 at 1.73 diameters, falls to a minimum near 1 at 2.5 diameters, and levels off. The hard disc null curve has the same shape but peaks at only 2.1, at the shorter separation of 1.57 diameters.}}%
\includegraphics[width=0.62\textwidth]{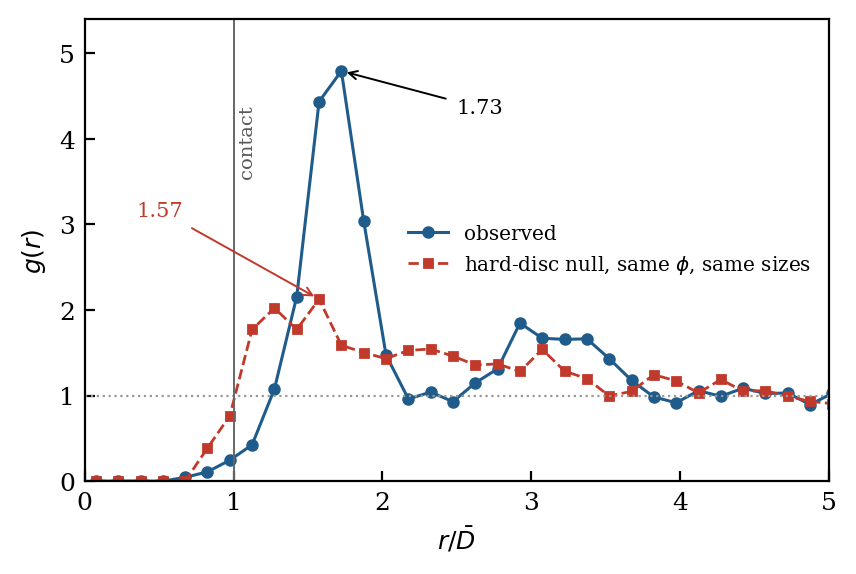}%
\EndAccSupp{}
\caption{Pair correlation function (blue) against a hard-disc
random-sequential-adsorption null at the same area fraction $\phi_{\rm nr} =
\num{0.17}$ and size distribution (red). The peak position is robust to
segmentation; its height is sensitive to the tail-normalisation window and
should not be treated as a measured quantity.}
\label{fig:gr}
\end{figure}

The pair correlation function (figure~\ref{fig:gr}) shows a hard exclusion
region below one diameter, a first coordination shell peaking at
$r/\bar{D} = \num{1.73}$ with $g = \num{4.8}$, a minimum near
\num{2.5} and a weak second shell. The hard-disc null at the same area fraction
and size distribution reaches only $g = \num{2.1}$, at the closer
separation $r/\bar{D} = \num{1.57}$. The observed shell is thus more
than twice as high as excluded volume alone produces, and sits further out.

\subsection{The order is liquid-like, not crystalline}
\label{sec:liquid}

\begin{figure}[t]
\centering\BeginAccSupp{method=escape,ActualText={Four panels on colony neighbour arrangement. Panel a, a micrograph overlaid with a triangular mesh joining neighbouring colonies, each colony coloured by how hexagonal its neighbourhood is, most values intermediate. Panel b, two overlapping histograms, observed centred at 0.46 and hard disc null at 0.39, both below the shaded hexatic band starting at 0.5. Panel c, bar chart of neighbour counts peaking at five rather than the six a crystal requires. Panel d, correlation of local hexagonal order falling from 0.7 to zero within 2.5 diameters, with exponential and power law fits.}}%
\includegraphics[width=\textwidth]{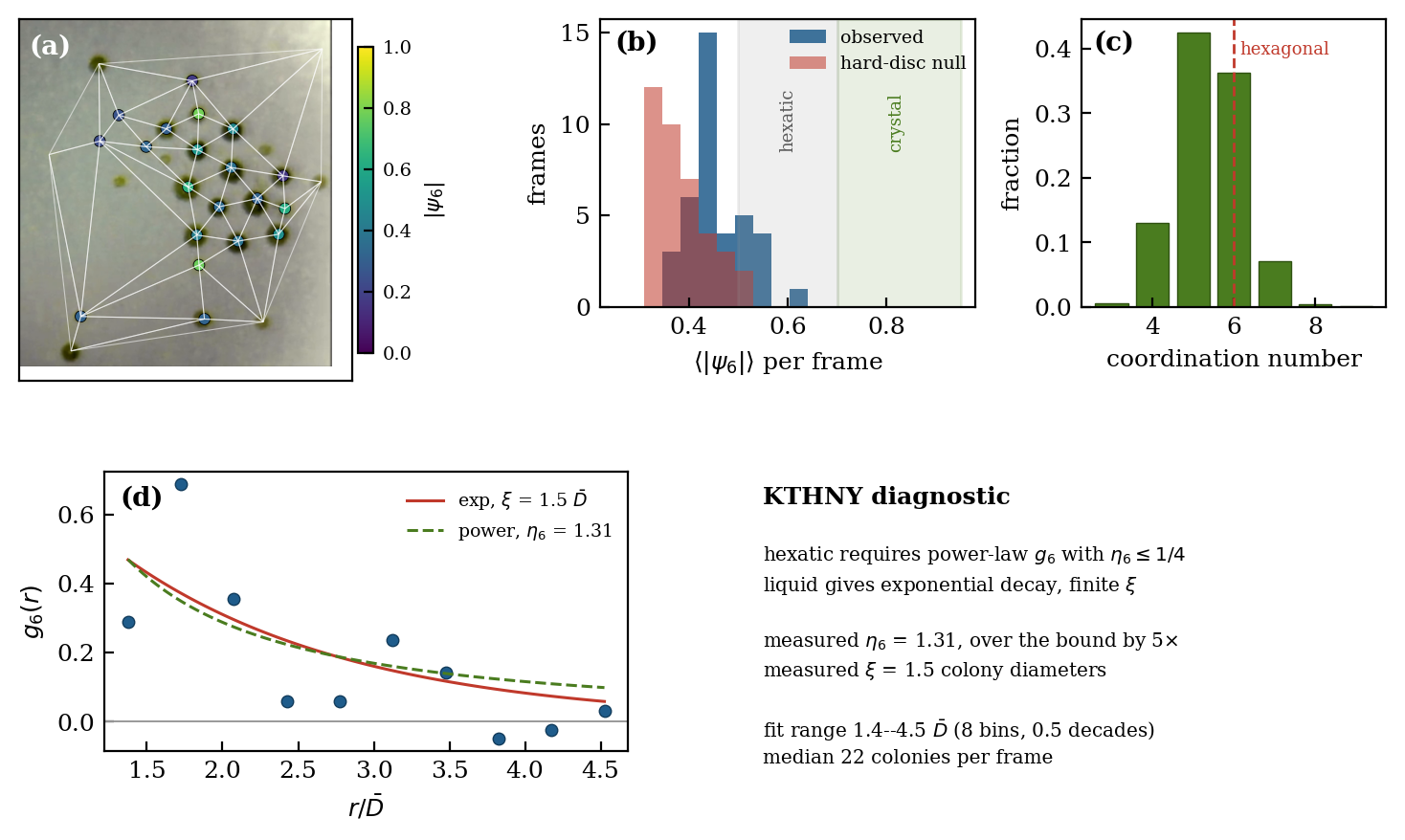}%
\EndAccSupp{}
\caption{(a) Delaunay triangulation of a representative frame, interior
colonies coloured by $|\psi_6|$. (b) Distribution of frame-averaged
$\langle|\psi_6|\rangle$, observed against the hard-disc null, with the
conventional hexatic and crystalline ranges shaded. (c) Delaunay coordination
numbers of interior colonies; the mode is \num{5}, not
\num{6}. (d) Orientational correlation function $g_6(r)$ with exponential and
power-law fits; the panel at right summarises the KTHNY diagnostic.}
\label{fig:psi6}
\end{figure}

A strong first shell in $g(r)$ establishes short-range order but does not by
itself distinguish a liquid from a crystal; what separates them is whether
orientational and translational order persist to large separation. Three
measurements say the arrangement is liquid-like (figure~\ref{fig:psi6}).

First, $g(r)$ decays to unity by $r/\bar{D} \approx 3.8$. The second shell is
present but broad, spanning $r/\bar{D} \approx 2.9$--$3.4$, rather than
resolving into the separate $\sqrt{3}$ and $2$ peaks a hexagonal lattice
produces. We note in passing that the first shell falls at
$r/\bar{D} = \num{1.73}$ and $\sqrt{3} = 1.732$; this is a coincidence
of expressing separations in colony diameters and is not evidence of a
hexagonal lattice constant.

Second, the local bond-orientational order parameter of
equation~\eqref{eq:psi} averages
$\langle|\psi_6|\rangle = \num{0.456} \pm \num{0.057}$ over
\num{38} frames and \num{574} interior colonies. For
reference, two-dimensional crystals give \numrange{0.7}{0.95}, a hexatic phase
\numrange{0.5}{0.7}, and a liquid \numrange{0.3}{0.45}. The measured value sits
at the top of the liquid range and short of hexatic. The hard-disc null gives
$\num{0.385} \pm \num{0.054}$, so the orientational order does exceed
random packing---consistent with the $g(r)$ result---but by only about
\num{0.07}. The square-symmetry control gives
$\langle|\psi_4|\rangle = \num{0.317}$, not far below $\psi_6$, which is
what a disordered arrangement rather than a six-fold symmetric one produces.

Third, Delaunay coordination numbers of interior colonies have mean
\num{5.38} and mode \num{5}, with only
\SI{36}{\percent} six-coordinated and a distribution running from
three to nine. A hexagonal crystal has essentially every interior particle
six-coordinated.

Fourth, and most directly, the KTHNY criteria are stated in terms of how
orientational correlations decay rather than how large they are. Equation
\eqref{eq:g6} gives a $g_6(r)$ that falls to the noise level by
$r/\bar{D} \approx 2.4$ (figure~\ref{fig:psi6}d). An exponential fit returns a
correlation length $\xi = \num{1.5}$ colony diameters; forcing a power law
instead returns $\eta_6 = \num{1.31}$. A hexatic phase requires power-law
decay with $\eta_6 \leq 1/4$, so the measured value exceeds the bound by a
factor of \num{5}. Short-ranged orientational correlations with a
finite $\xi$ are the signature of a liquid.

Two limits on that inference should be stated. The cluster spans only about
five colony diameters, so the fit covers $r/\bar{D} =
\numrange{1.4}{4.5}$, \num{8} bins and roughly half a
decade; discriminating exponential from power-law decay on goodness of fit
alone is therefore weak, and it is the magnitude of $\eta_6$ relative to $1/4$,
not the choice of functional form, that carries the argument. We also did not
attempt the companion translational correlation $g_G(r)$: with a median of
\num{22} colonies per frame there is no resolvable structure-factor
peak from which to take a reciprocal lattice vector, and imposing one by hand
produces an artefact of the imposed wavevector rather than a measurement.
Finally, KTHNY is an equilibrium theory for passive two-dimensional systems. We
use its criteria as a diagnostic, and do not claim that the theory applies to
an active suspension.

We therefore describe the arrangement as a two-dimensional liquid with
unusually strong short-range order, and specifically not as a crystal. The
result of interest is that this liquid is more ordered than hard discs at the
same area fraction, not that it is close to crystalline.

\subsection{Single-colony dynamics}

\begin{figure}[t]
\centering\BeginAccSupp{method=escape,ActualText={Three panels on colony movement. Panel a, a stem plot of brightness difference between consecutive video frames, alternating strictly between 0.04 and 1.5, revealing that every second frame is a duplicate. Panel b, log-log plot of mean squared displacement against time lag, slope near 2 at short times bending to near 1 beyond about 2 seconds. Panel c, velocity autocorrelation falling from 1.0 through a plateau near 0.24 and crossing zero at 1.6 seconds.}}%
\includegraphics[width=\textwidth]{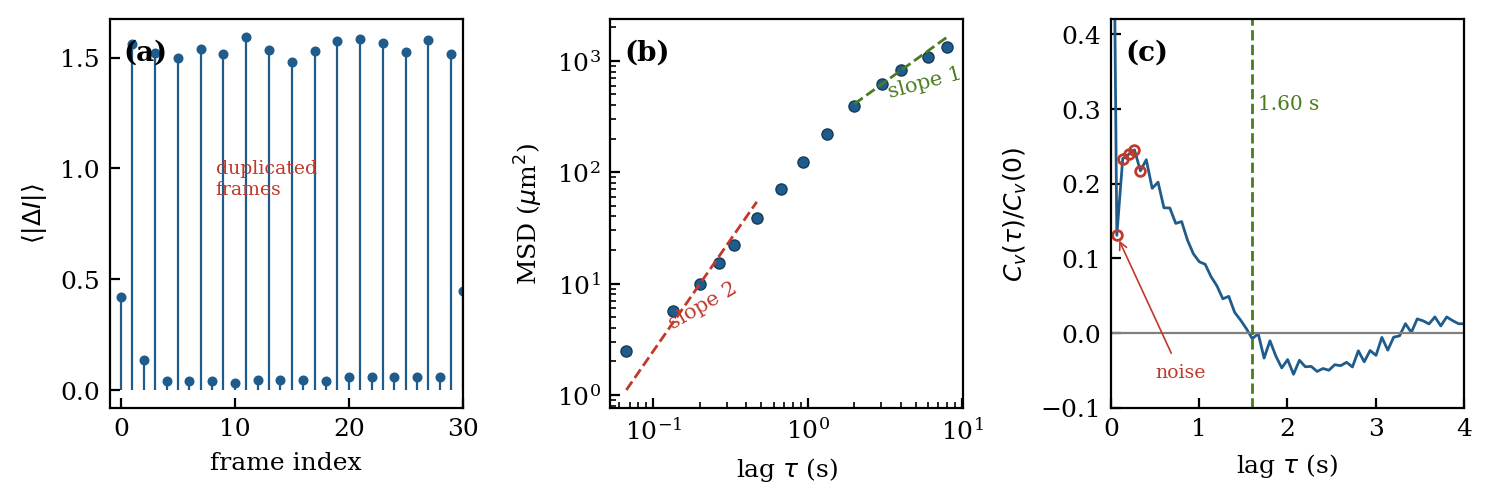}%
\EndAccSupp{}
\caption{(a) Frame-difference diagnostic showing the period-two alternation
that reveals frame duplication in the native-rate clip. (b) Mean-squared
displacement, \num{24} tracks, with ballistic and diffusive guides.
(c) Normalised velocity autocorrelation; open circles mark the first lags,
where the depression is a measurement-noise artefact.}
\label{fig:dyn}
\end{figure}

From \num{24} continuously tracked colonies the mean-squared
displacement (figure~\ref{fig:dyn}b) is superdiffusive at short lag, local
logarithmic slope \num{1.4}, crossing to \num{0.8} beyond about
\SI{2}{\second}: a persistent random walk. The short-lag exponent falls below
the ballistic value of \num{2} because additive centroid noise contributes a
lag-independent offset.

The velocity autocorrelation (figure~\ref{fig:dyn}c) crosses zero at
\SI{1.6}{\second}, consistent with the MSD crossover; under the alternative
segmentation the crossing falls at
\SIrange{1.4}{1.7}{\second}. The first lag is depressed by the
anti-correlation that differencing white centroid noise produces, and is
excluded. A weak negative lobe beyond the crossing indicates trajectories that
curve back on themselves.

Taking $v$ from the ballistic regime gives
\SIrange{13}{17}{\micro\metre\per\second} across methods, persistence
length $v\tau \approx \SI{21}{\micro\metre}$ and $D_{\rm eff} \approx
\SI{41}{\micro\metre\squared\per\second}$. The Reynolds number is
$\sim 9\times 10^{-4}$. The speed is an order of magnitude below free-swimming
volvocine values~\cite{solari2008}, as expected for colonies swimming into a
rigid wall.

\subsection{Lifetime and breakup}

The arrangement is not indefinitely stable. Taking the stable-phase mean plus
three standard deviations as a threshold, the first sustained excursion occurs
at \SI{127}{\minute}. Beyond \SI{105}{\minute} the median
nearest-neighbour ratio rises to \num{2.25}, the colony count falls from
\num{30} to \num{19}, and the area fraction falls from
\num{0.104} to \num{0.043}. The breakup is therefore a dilution
of the layer rather than a collapse into contact. We did not determine the
cause; oxygen depletion, nutrient exhaustion, changing illumination and loss of
motility in an ageing sample are all uncontrolled.

\subsection{Colony morphology}
\label{sec:morphology}

The size distribution is bimodal, with a dominant mode near
\SI{71}{\micro\metre} and a smaller mode consistent with released
juveniles or free unicells, pooled over \num{2193} detections. Bimodality
is expected for a coenobium reproducing by autocolony formation, and is
plausibly sharpened by synchronisation to the light:dark cycle during
enrichment.

\begin{figure}[t]
\centering\BeginAccSupp{method=escape,ActualText={Two panels on colony internal structure. Panel a, a magnified image of a single colony, a dark green disc about 70 micrometres across with a slightly lighter centre and no resolvable individual cells. Panel b, four curves of normalised brightness against distance from the colony centre divided by radius, all rising from a shallow dip at the centre to a maximum between 0.3 and 0.5 of the radius before falling at the edge. The central dip is only 1 to 4 per cent deep.}}%
\includegraphics[width=0.92\textwidth]{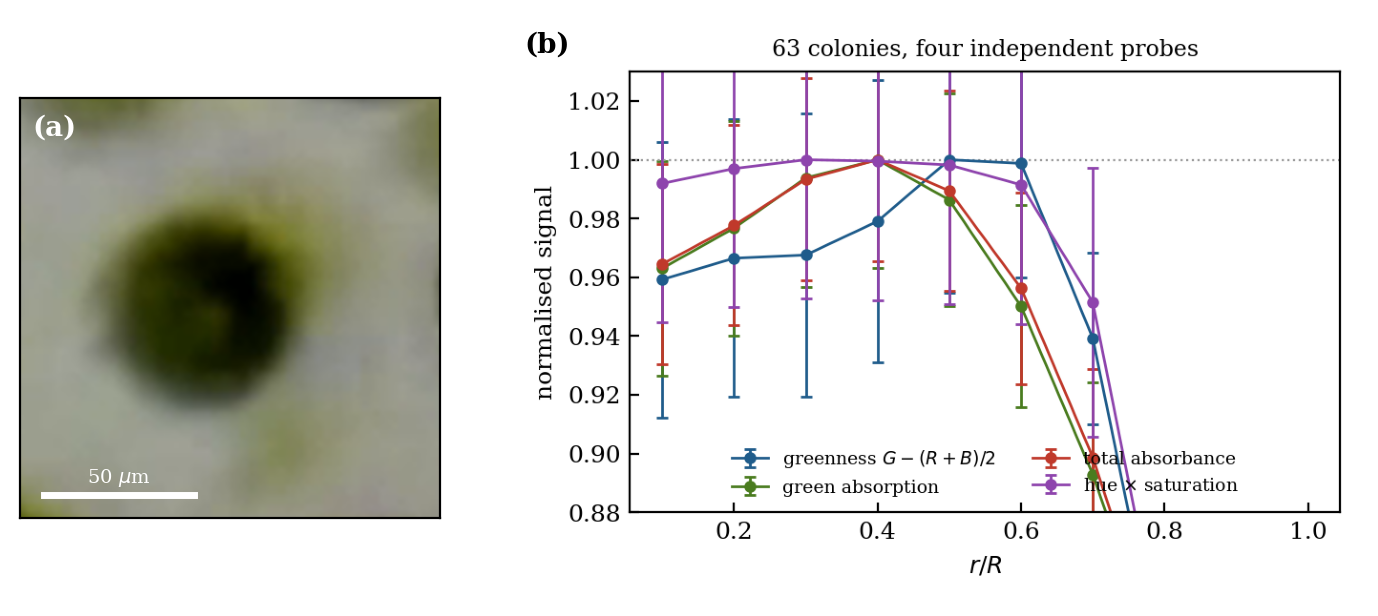}%
\EndAccSupp{}
\caption{(a) A single colony, digitally magnified; individual cells are not
resolved. (b) Azimuthally averaged signal against normalised radius for
\num{63} colonies under four optical probes, each normalised to its own
maximum. All place the maximum off centre; the central deficit is only
\SIrange{1}{4}{\percent} deep.}
\label{fig:hollow}
\end{figure}

Azimuthally averaged signal profiles across \num{63} colonies under
four independent optical probes place the intensity maximum off centre, at
$r/R = 0.3$--$0.5$ (figure~\ref{fig:hollow}), the direction a hollow coenobium
predicts. The magnitude is small---the central deficit is between
\SI{1}{\percent} and \SI{4}{\percent} depending on probe---and
on images of this quality would not have been sufficient to establish the
internal structure on its own. It is consistent with the morphology confirmed
in \S\ref{sec:identification} rather than evidence for it, and it illustrates
the limit of what the low-cost instrument can resolve.

The dominant colony diameter of \SI{71}{\micro\metre} sits inside the
\emph{Eudorina} range and well outside that of \emph{Volvox}, which begins near
\SI{300}{\micro\metre}. Because the calibration is nominal rather than
metrological (\S\ref{sec:optics}), this agreement is worth stating in the other
direction as well: the independently confirmed identification is evidence that
the assumed length scale is approximately correct, and it is the only
independent check on it that we have.

\subsection{Robustness to method}
\label{sec:robust}

Recomputing with the alternative segmentation, three results are stable: the
$g(r)$ first-shell position moves from \num{1.73} to \num{1.72}; the
mean colony diameter moves from \SI{71}{\micro\metre} to
\SI{68}{\micro\metre}; the size distribution remains bimodal. Three are not:
the nearest-neighbour median rises to
\numrange{1.91}{2.07}, because the alternative
segmentation detects fewer small colonies and a missed colony lengthens its
neighbours' separations; the swimming speed rises to
\SI{17}{\micro\metre\per\second}; and the excess kurtosis of the
log-diameter distribution changes sign, so that statistic should not be used as
evidence of bimodality even though the bimodality itself survives.

We therefore treat the $g(r)$ peak position as the primary structural result
and the nearest-neighbour median as secondary, quoted with its method spread.

\begin{table}[t]
\centering\small
\begin{tabular}{@{}lll@{}}
\toprule
Quantity & Value & Note \\
\midrule
Colony diameter (dominant mode) & \SI{71}{\micro\metre} & nominal calibration \\
$g(r)$ first-shell position & $r/\bar{D} = 1.73$ & \num{1.72} alt.\ method \\
$g(r)$ first-shell height & $4.8$ & null: $2.1$ at $1.57$ \\
Area fraction & $\phi_{\rm nr}=\num{0.17}$, $\phi_{\rm tl}=\num{0.104}$ & convex hull, two sequences \\
Nearest-neighbour separation & $1.80 \pm 0.13$ diameters & \numrange{1.91}{2.07} alt. \\
Bond-orientational order $\langle|\psi_6|\rangle$ & $0.456 \pm 0.057$ & null: $0.385 \pm 0.054$ \\
Six-coordinated interior colonies & \SI{36}{\percent} & crystal: $\approx$\SI{100}{\percent} \\
$g_6$ correlation length $\xi$ & \num{1.5} diameters & exponential fit \\
$g_6$ power-law exponent $\eta_6$ & \num{1.31} & hexatic bound: $1/4$ \\
Reorientation time $\tau$ & \SIrange{1.4}{1.7}{\second} & MSD and VACF agree \\
Swimming speed $v$ & \SIrange{13}{17}{\micro\metre\per\second} & across methods \\
Persistence length $v\tau$ & \SI{21}{\micro\metre} & derived \\
$D_{\rm eff}$ & \SI{41}{\micro\metre\squared\per\second} & diffusive MSD \\
Reynolds number & $\sim 9\times 10^{-4}$ & $vD/\nu$ \\
Stable phase & \SI{105}{\minute} & breakup onset \SI{127}{\minute} \\
Record length & \SI{153}{\minute} & elapsed \\
\bottomrule
\end{tabular}
\caption{Summary. Dimensionless quantities are independent of the calibration
uncertainty; those in microns are not. Every entry is read from the deposited
data by the script that typesets this table.}
\label{tab:res}
\end{table}

\section{Discussion}

A population of swimming colonies held against a rigid boundary adopts a
two-dimensional arrangement with liquid-like short-range order, a preferred
separation near two diameters, and a lifetime of order two hours. The hard-disc
comparison rules out the trivial explanation: at the measured area fraction,
excluded volume produces a first shell less than half as high and at a smaller
separation, so something is holding the colonies apart at a specific distance.

The order is liquid-like rather than crystalline
(\S\ref{sec:liquid}), which is the more interesting result: a liquid that is
measurably more ordered than hard discs at the same density implies an
interaction, whereas a crystal at this area fraction would more likely indicate
jamming.

The obvious candidate is the surface-mediated hydrodynamic interaction
identified for \emph{Volvox} by Drescher \emph{et al.}~\cite{drescher2009}, in
which a spinning bottom-heavy swimmer near a no-slip wall generates a flow that
couples to its neighbours. Our chamber has the rigid boundary that analysis
assumes. However, our colonies are several times smaller than \emph{V.\
carteri}, and we have not demonstrated that they spin, nor measured the flow
field, so this remains a hypothesis rather than a conclusion.

Two limitations bound what can be drawn from this work. The length
calibration is nominal, so all micron-valued quantities carry an uncertainty we
cannot bound rigorously; this does not affect the dimensionless results, which
are the substance of the paper. And this is a single population observed once:
we have no replication, no density series, and no free-surface control. A density series
would distinguish a pair interaction, which should give a
density-independent preferred separation, from a collective effect, which
should not. A free-surface repeat would test the boundary-condition dependence
directly, since the hydrodynamic image system of a spinning swimmer differs
between a no-slip wall and a stress-free surface.

We report the measurement nonetheless because the spatial statistics of
boundary-accumulated algal layers appear not to have been quantified for these
smaller colonial genera, and because the null-model comparison and the measured
lifetime are both directly reusable by anyone repeating the experiment.

\subsection*{Cost as a feature}
\label{sec:cost}

The entire apparatus is a handheld USB microscope, a clear plastic coin capsule
and a sample of pond water. There is no laser, no high-speed camera, no
microfabrication, no temperature control, no optical table and no consumable
beyond the capsule itself. Total outlay is a small fraction of a single
research-grade objective, and the analysis runs on a laptop using
open-source libraries.

Active matter is normally studied with equipment that concentrates the field in
well-funded laboratories: microfabricated channels, confocal or high-speed
imaging, precision thermostatting. The measurements reported here---a pair
correlation function, a bond-orientational order parameter, an orientational
correlation length, a mean-squared displacement---are standard soft-matter
observables, and none of them required any of that. The cost of using an
inexpensive apparatus is metrological: an uncalibrated length scale and an unmeasurable depth
of field, limitations we have set out in detail in \S\ref{sec:optics}. What it
does not cost is the physics, because every conclusion we draw is
dimensionless and therefore independent of the calibration.

A primary motivation in our development of a low-cost apparatus was to bring
it within reach of undergraduate teaching laboratories and of research groups in
institutions with limited equipment budgets, including those in low- and
middle-income countries where the organisms themselves are abundant.

\section*{Ethics}

This work involved environmental sampling of freshwater microorganisms and
required no ethical approval.

\section*{Data accessibility}

All raw video, derived data, and Python and MATLAB analysis code, including the
independent-verification script used in \S\ref{sec:robust}, are available at
\href{https://doi.org/10.7267/0g354q328}
{\texttt{doi:10.7267/0g354q328}}, in ScholarsArchive@OSU, the institutional
repository of Oregon State University
(\url{https://ir.library.oregonstate.edu/concern/datasets/0g354q328}). The
deposit includes all three video sequences: the two analysed
here and the record of sample collection. The collection footage and the
time-lapse sequence are additionally available online at
Refs.~\cite{video1,video2}.

\section*{Declaration of AI use}

The author used Claude (Anthropic) to edit the manuscript for clarity and to
assist in debugging the analysis code. The author reviewed and validated all
AI-assisted output and takes full responsibility for the content.

\section*{Conflict of interest declaration}

The author declares no competing interests.

\section*{Funding}

No external funding was received for this work.

\end{document}